\documentclass[natbib]{svjour3}    
\journalname{Celestial Mechanics and Dynamical Astronomy}

\smartqed  

\usepackage{amsmath,amssymb}
\usepackage{graphicx}
\usepackage{mathptmx}

\usepackage{latexsym}

\newcommand{\Z}{\mathbb{Z}}

\newcommand{\RN}{\mathbb{R}}

\def\E{\mathcal E}
\def\C{\mathcal C}

\def\T{\mathcal T}
\let\e=\epsilon
\let\d=\delta

\newcommand{\be}{\begin{equation}}
\newcommand{\ee}{\end{equation}}
\newcommand{\ba}{\begin{eqnarray}}
\newcommand{\ea}{\end{eqnarray}}
\newcommand\beq[1]{ \begin{equation}\label{#1} }
\newcommand{\eeq}{ \end{equation} }

\newcommand\beqa[1]{ \begin{eqnarray} \label{#1}}
\newcommand{\eeqa}{ \end{eqnarray} }
\newcommand{\beqano}{ \begin{eqnarray*} }
\newcommand{\eeqano}{ \end{eqnarray*} }
\newcommand\equ[1]{{\rm (\ref{#1})}}

\def\bbm[#1]{\mbox{\boldmath $#1$}}

\begin{document}

\title{Structure of the center manifold of the $L_1,L_2$ collinear libration points in the restricted three-body problem}


\author{Giuseppe Pucacco
}


\institute{              G. Pucacco \at Dipartimento di Fisica and INFN -- Sezione di Roma II,
Universit\`a di Roma ``Tor Vergata", \\
Via della Ricerca Scientifica, 1 - 00133 Roma\\
\email{pucacco@roma2.infn.it} }
\maketitle

\begin{abstract}
We present a global analysis of the center manifold of the collinear points in the circular restricted three-body problem. The phase-space structure is provided by a family of resonant 2-DOF Hamiltonian normal forms. The near 1:1 commensurability leads to the construction of a detuned Birkhoff-Gustavson normal form. The  bifurcation sequences of the main orbit families are investigated by a geometric theory based on the reduction of the symmetries of the normal form, invariant under spatial mirror symmetries and time reversion. This global picture applies to any values of the mass parameter.


\end{abstract}

\section{Introduction}
\label{intro}
Motion around the collinear points of the spatial restricted three-body system is a classical and rich problem in  Celestial Mechanics. Historically the study started with the investigation of orbit families with particular emphasis on the `halo' orbits, using at first pioneering numerical methods \citep{FK73,Henon1} and subsequently perturbation theory methods like the Poincar\'e-Lindstedt approach \citep{richardson,JM}. Whereas the description of these orbits became satisfactory enough to be usefully exploited as a basis for space missions \citep{Howell84,GJMS}, a deep and general understanding of the general orbit structure has been obtained only quite recently with works devoted to the systematic construction of the invariant manifolds around the equilibria \citep{GM,CCP,DGR,GL18,WW}. 

Here we want to describe the main features of the center manifold of $L_1$ and $L_2$, namely the subspace remaining after the elimination of the stable/unstable submanifolds at each energy level. The motivation comes from the necessity of `seeing' their orbit structure, using an expression due to  \citet{GU18} and has also been inspired by the approach adopted by \citet{Lara17} to treat the case of the Hill problem. The structure of the center manifold can most effectively be obtained from a normal form \citep{CPS,CCP} which captures both the hyperbolic dynamics {\em normal to} and the elliptic dynamics {\em on} the center manifold. The normal form is constructed from the hyperbolic-elliptic-elliptic linear term in which the two unperturbed elliptic frequencies are very close to resonance and the hyperbolic dynamics are encoded in the conservation of a formal integral of motion. After the center manifold reduction, performed by choosing initial data such that this integral vanishes, what remains is a standard `detuned' resonant normal form \citep{Henrard,Vf}. Here we show that already the first-order truncation of the normal form series is generic in such a way to capture the main features of the phase-space around the collinear equilibria \citep{MP14,MP16}. Higher-order terms improve only quantitatively the analytical predictions. 

A most effective description is achieved by means of the invariants of the isotropic oscillator \citep{CB,HH18} which plays the role of resonant unperturbed system. Then, a singular reduction process of the normal form \citep{KE} does the rest of the job. Fitting the coefficients of the normal form in order to get a model valid for arbitrary values of the mass parameter, as done in \citet{CCP}, we get a `universal' description of the main bifurcations associated to the resonance. 

We stress here the global structural stability of this family of models: their features do not change under small perturbations of the parameters. The proof of this statement can most easily be obtained by exploiting the geometric reduction approach. The main ideas on which this method grounds \citep{CR,han07} are highlighted hereafter in order to have a self-contained setting. The resulting tool-box is a set of actions and mappings from orbits on the phase-space to points on a reduced space. In particular, periodic orbits are mapped to critical points. Their cardinality and nature provide the backbone of the family of models. Changes produced by varying the energy level are associated with bifurcations and stability transitions. The normal modes of the system (the `Lyapunov orbits') exist at arbitrary energy. At specific energy thresholds, bifurcation of periodic orbits in `generic position' \citep{SV} occur, producing the halo (oscillations in quadrature of the resonating modes) and, possibly, the `anti-halo' periodic orbits (oscillations in phase). Their existence organise the whole structure of the family of models.

A remarkable feature of the family, valid for arbitrary $\mu$, is that it is quite close to the degenerate case proper of the detuned H\`enon-Heiles system \citep{HS}, but always distinct from it. This produces a relatively low value for the energy threshold at which halo orbits appear and, at much higher energy levels, a close pair of energy values at which the `bridge' of the anti-halo families bifurcate from the planar Lyapunov and annihilate on the vertical one. The momentum map can be easily computed \citep{PHD} providing a clear view of the invariant structure of periodic orbits and invariant tori. Implications on the original system can be deduced by inverting the normalising transformation.

The plan of the paper is as follows: in Section 2 we compute the resonant normal form necessary to capture the bifurcations associated to the synchronous resonance; in Section 3 we illustrate the geometric reduction provided by mapping the phase space to the space of the invariants of the isotropic oscillator; in Section 4 we further reduce the symmetries of the normal form and analyse the bifurcation scenario; in Section 5 we discuss the results and hints for future investigations; in Section 6 we give our conclusions.

\section{The normal form around the collinear equilibria}
\label{NFS}

The model usually adopted to describe the dynamics around the collinear equilibria of the circular restricted three-body problem is given by the following Hamiltonian:
\beq{ham1}
H(p_x,p_y,p_z,x,y,z)=\frac{1}{2}\left( p_x^2+p_y^2+p_z^2\right)+yp_x-xp_y- \sum_{n\ge2}c_n(\mu) T_n (x,y,z) \ .
\eeq
We adopt the general conventions used e.g. in \citet{CPS} to which we refer for the details. Here we recall that the coefficients $c_n(\mu)$, different for each equilibrium, are uniquely expressed in terms of the mass parameter $0 < \mu \le 1/2$ as
\beq{cn}
c_n(\mu)=\frac{(\pm1)^n}{\gamma^3}\left(\mu+(\mp1)^n\frac{(1-\mu)\gamma^{n+1}}{(1 \mp \gamma)^{n+1}}\right) 
\eeq
where $\gamma$,  the distance of the collinear equilibrium to the closest primary, is the solution of the fifth-order Euler's equations:
\beq{eulerq}
\gamma^5 \mp (3-\mu)\gamma^4+(3-2\mu)\gamma^3-\mu\gamma^2 \pm 2\mu \gamma-\mu=0.  
\eeq
In eqs.(\ref{cn},\ref{eulerq}) the upper sign refers to $L_1$, the lower one to $L_2$. 
The polynomials $T_n$,  related to the Legendre polynomials by 
$$T_n(x,y,z)\equiv r^n {\mathcal P}_n\left(\frac{x}{r}\right), \quad r=\sqrt{x^2+y^2+z^2},$$ 
are recursively defined by means of
$$
T_0=1\ , \quad T_1=x\ , \quad  ... \, , \quad T_n=\frac{2n-1}{n}xT_{n-1}-\frac{n-1}{n} \left(x^2+y^2+z^2\right) T_{n-2}\ .
$$
Hamiltonian \eqref{ham1} is endowed with the symmetry given by its invariance under the reflection
\beq{zrefl}
z \rightarrow -z,\eeq
the mirror symmetry with respect to the synodic plane which plays an important role in the following.

\subsection{Diagonalisation of the quadratic part}\label{sec:diagonalization}
\begin{figure}[h]
\center
\includegraphics[height=7cm]{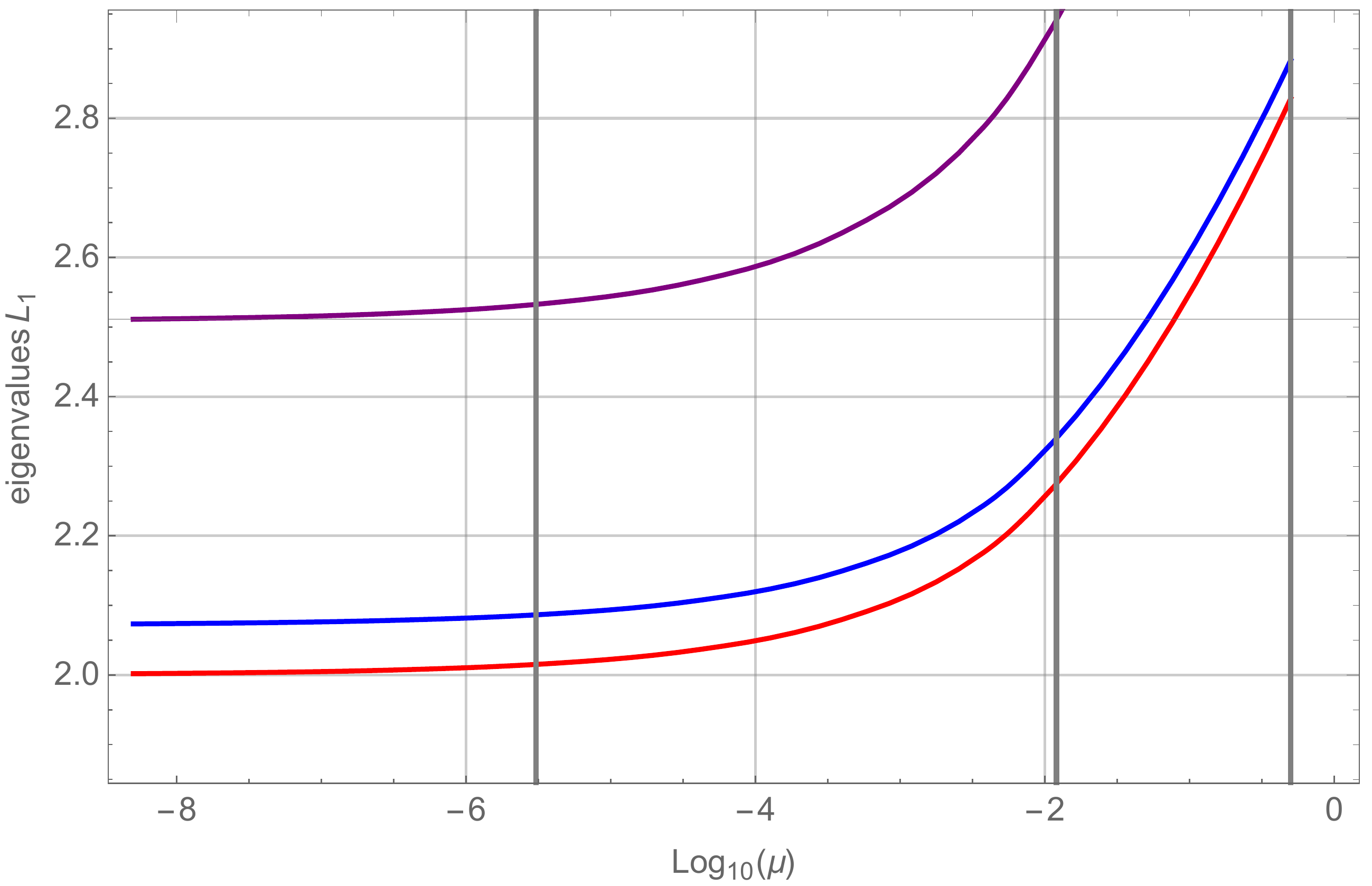}\\
\includegraphics[height=7cm]{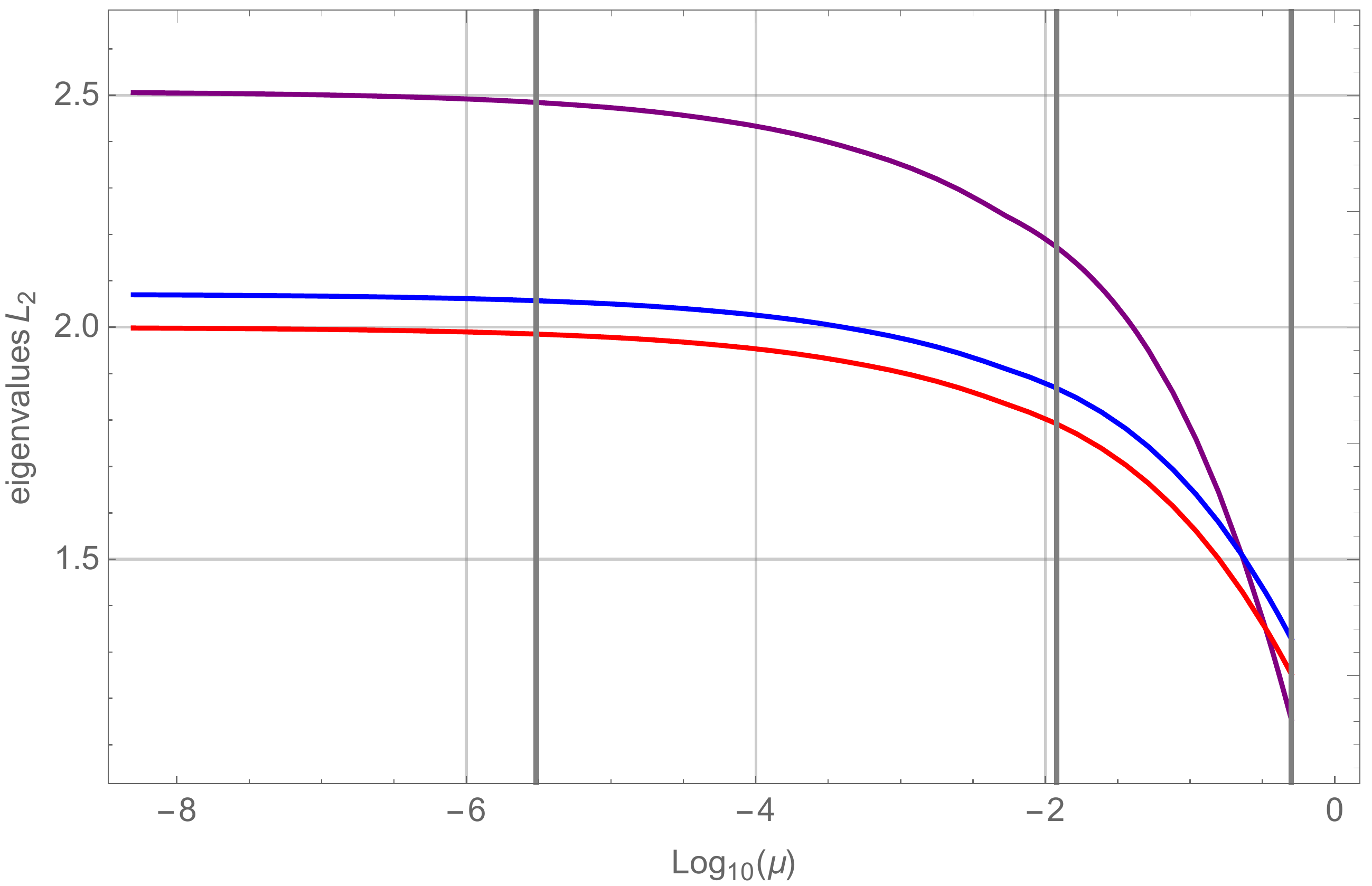}\\
\caption{ The eigenvalues $\lambda_x$ (purple), $\omega_y$ (blue) and $\omega_z$ (red) in terms of the mass ratio, for $L_1$ (upper panel) and $L_2$ (lower panel). Grey lines correspond to the three `reference' cases: Earth-Sun ($\mu=3.04\times10^{-6}$), Earth-Moon ($\mu=0.012$) and equal masses ($\mu=1/2$). 
}\label{FR1}
\end{figure}

Linearising \equ{ham1} around each equilibrium point, the quadratic part of the Hamiltonian can be written as:
\beq{H2}
H_0(p_x,p_y,p_z,x,y,z)=\frac{1}{2}\left(p_x^2+p^2_y\right)+yp_x-xp_y+\frac{c_2}{2}\left(y^2-2x^2\right)+\frac12\left(p_z^2+ c_2 z^2\right) \ ,
\eeq
where  the explicit dependence on $\mu$ has been dropped from the coefficient. It appears that
\beq{omegaz}
\sqrt{c_2} \doteq \omega_z
\eeq 
can be interpreted as the `vertical' linear frequency.  The other two eigenvalues of the diagonalising matrix \citep{JM} are 
\beq{E2}
\lambda_x=\frac{\sqrt{c_2-2+\sqrt{9c_2^2-8c_2}}}{\sqrt{2}}\ , \qquad
\omega_y=\frac{\sqrt{2-c_2+\sqrt{9c_2^2-8c_2}}}{\sqrt{2}}\ .
\eeq
 For further reference we plot in Fig. \ref{FR1} the three eigenvalues as functions of $\mu$.  
The quadratic part of the Hamiltonian can be put into the diagonal form
\beq{H2D}
H_0(p,q)=\lambda_x q_1p_1+i\omega_y q_2p_2+i\omega_z q_3p_3\ ,
\eeq
where $(p,q)=(p_1,p_2,p_3,q_1,q_2,q_3)$ denote the new diagonalising variables. It appears clear the `saddle $\times$ center $\times$ center' nature of the linearised motion, with $\lambda_x$ giving the hyperbolic rate along the directions tangent  to the stable and unstable hyper-tubes \citep{GL18} and with $\omega_y$ playing the role of `horizontal' linear elliptic frequency. 

\subsection{Resonant normalisation}\label{norm:form}
In the diagonalising variables the Hamiltonian is the series 
\beq{ham1d}
H(p,q)=\sum_{n\ge 0}H_n(p,q)\ ,
\eeq
where $H_0$ is given by \eqref{H2D} and $H_n$ are homogeneous polynomial in $(p,q)$ of degree $n+2$. The construction of the normal form is performed with a twofold objective: 

\noindent
1. to get an expression in which is easy to `kill' the evolution along the stable-unstable manifolds; 

\noindent
2. to include the passage through the synchronous resonance. 

\begin{figure}[h]
\center
\includegraphics[height=7cm]{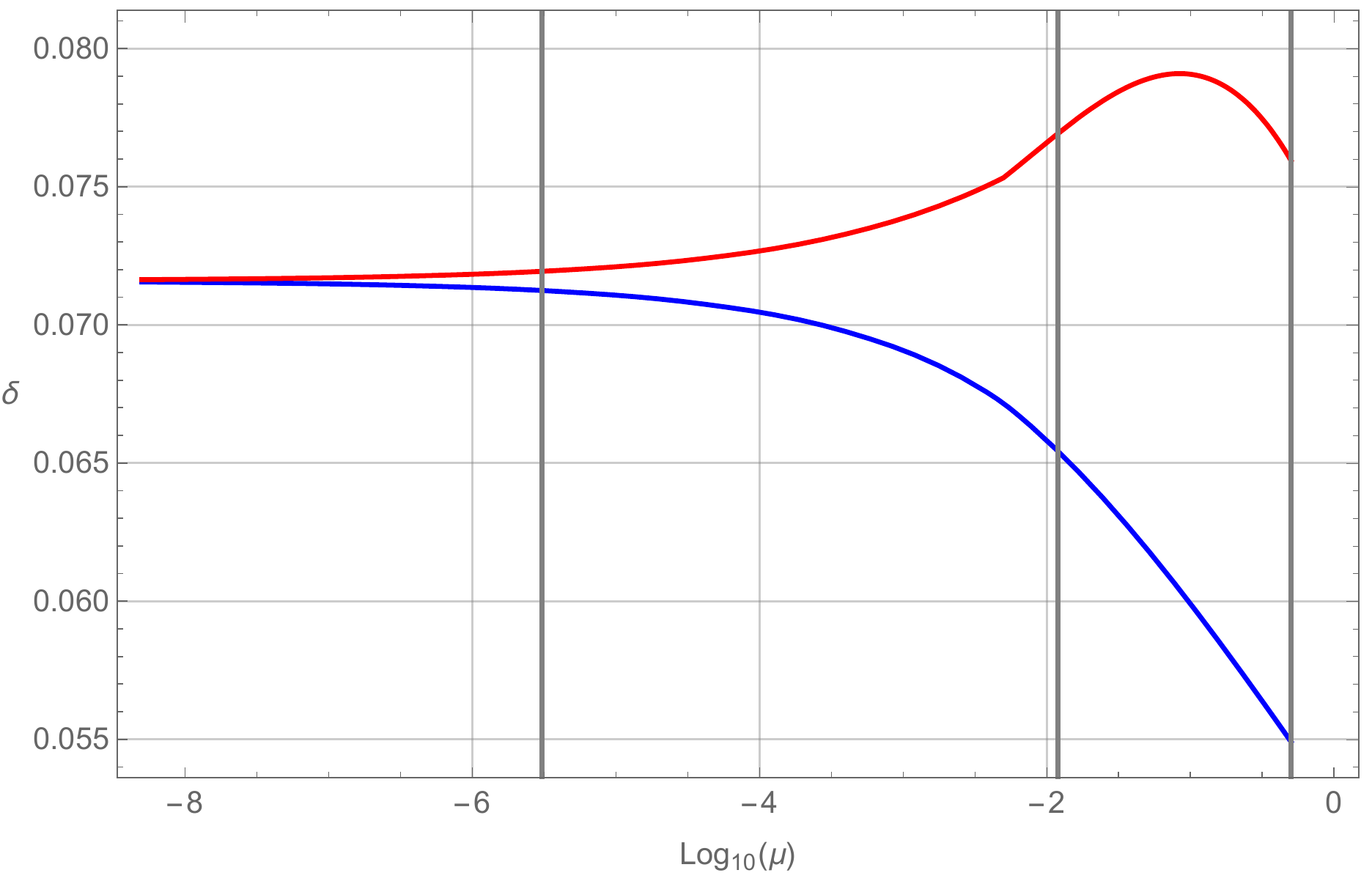}
\caption{ The detuning parameter \eqref{detd}: $L_1$ (blue) and $L_2$ (red). 
}\label{FR2}
\end{figure}

 In fact, from Fig. \ref{FR1}, for any $0 < \mu \le 1/2$, we see that  the two elliptic frequencies are very close to each other  and,  introducing the `detuning'
\beq{detd}
\delta \doteq \omega_y-\omega_z\ ,
\eeq
the following series expansions 
\beq{detser}
\delta=\delta_0\pm\delta_1\mu^{1/3}\pm\delta_2\mu^{2/3}+\delta_3^{(\pm)}\mu+{\rm O}(\mu^{4/3})
\eeq
have been obtained by \citet{CCP}\footnote{The coefficients are listed in two Boxes at the end of \citet{CCP}.} with 
$$\delta_0=-2 + \sqrt{-1 + 2 \sqrt{7}}\simeq0.0716 \ .$$ 
In \eqref{detser}, the upper sign refers to $L_2$, the lower one to $L_1$. These give a quite accurate approximation for mass parameters up to $10^{-2}$ and, in the most extreme case of $\mu=1/2$, give $\delta_{L_1}=0.0536$ and $\delta_{L_2}=0.0760$, while the exact values are $\delta_{L_1}=0.05492$ and  $\delta_{L_2}=0.07596$ (see   Fig. \ref{FR2}, from which we also see that $\delta$ is
monotonically decreasing for $L_1$  $\forall \mu \in (0,1/2]$, while for $L_2$ first increases, until
$\mu = 0.21728$ where it reaches the value 0.08128, and then decreases). 

It is then natural to assume the linear system to be close to the 1:1 resonance and construct a normal form which incorporates the corresponding resonant terms. Therefore, we look for a canonical transformation, $(p,q) \longrightarrow (P,Q)$,
which  formally  conjugates \equ{ham1d} to the Hamiltonian function
\beqa{zero}
K(P,Q)&=&\lambda_x Q_1P_1+i\omega_z ( Q_2P_2+  Q_3P_3)+i\delta Q_2P_2 + \sum_{n=1}^N K_{2n}(Q_1P_1,P_2,P_3,Q_2,Q_3)+R_{N+1}\ ,\nonumber\\
\eeqa
where the homogeneous polynomials $K_{2n}$, $n=1,...,N$, of degree $2(n+1)$, are in \sl normal form \rm with respect to the
resonant quadratic part
\beq{hqr}
H_0^{(r)}\doteq \lambda_x Q_1P_1+i\omega_z ( Q_2P_2+  Q_3P_3) \ .
\eeq
The terms in the normal form are of even degree in view of the combined effect of the reflection symmetry \eqref{zrefl} and of the commutation condition with \eqref{hqr}. This produces a function endowed with extra symmetries: invariance under time reversion (equivalent to the exchange $P_1\leftrightarrow Q_1$) and with respect to the $\Z_2\times\Z_2\times\Z_2$ group generated by
\beqa{spatial_symmetry}
\mathcal S_k&:&(P_k,Q_k)\rightarrow(-P_k,-Q_k), \quad \forall \ k=1,2,3.
\eeqa
In producing these symmetries, information is not lost but is encoded in the generating functions of the normalising transformation: for a general discussion of this and other related phenomena we refer to \citet{tv} and \citet{SV}. In \eqref{zero}, the detuning term is effectively considered to be of higher order and $R_{N+1}(P,Q)$ is a remainder function of degree $N+3$. Explicitly, the construction is implemented with the Lie method \citep{GG78,CG} and, for the purposes of the present work, is performed up to the first order at which resonant terms appear in the normal form, simply $N=1$. 

We remark that this method allows us to simultaneously perform the central manifold reduction {\it and} the analytic treatment of the synchronous bifurcations: as a matter of fact, the truncated normal form is an {\it integrable} 3-DOF system, in view of the presence of two additional formal integrals of motion, namely 
\beq{first_int}
J_x \doteq Q_1P_1
\eeq
and 
\beq{sec_int}
\E \doteq i( Q_2P_2 +  Q_3P_3 ) \ .
\eeq
For this integrable system all information about periodic and quasi-periodic motion can be retrieved. Previous works focused primarily on using Lie series only to compute the center manifold \citep{JM}. In fact, \citet{Lara17}, in his approach to study the center manifold of the Hill problem, has been forced to perform a further averaging with respect to the mean anomaly to get his final normal form. On the other hand, the Hamiltonian introduced by \citet{CPS},
\beqa{zerodue}
K(P,Q)&=&H_0^{(r)}+i\delta Q_2P_2 + K_{2}(Q_1P_1,P_2,P_3,Q_2,Q_3) \ ,
\eeqa
is the tool which incorporates both features stated in points 1. and 2. above.

\subsection{Center manifold reduction}\label{sec:center}
Introducing action-angle variables according to
\begin{equation}
\begin{cases}
Q_1=\sqrt{J_x}e^{\theta_x} \nonumber\\
Q_2=\sqrt{J_y}(\sin \theta_y -i\cos \theta_y)=-i\sqrt{J_y}e^{i\theta_y} \nonumber\\
Q_3=\sqrt{J_z}(\sin \theta_z -i\cos \theta_z)=-i\sqrt{J_z}e^{i\theta_z} \nonumber\\
P_1=\sqrt{J_x}e^{-\theta_x} \nonumber\\
P_2=\sqrt{J_y}(\cos \theta_y -i\sin \theta_y)=\sqrt{J_y}e^{-i\theta_y}  \nonumber\\
P_3=\sqrt{J_z}(\cos \theta_z -i\sin \theta_z)=\sqrt{J_z}e^{-i\theta_z}\ ,\nonumber
\end{cases}
\end{equation}
looking only at the terms in the truncated series \eqref{zerodue}, it appears that, as said above,  $J_x=Q_1P_1$ is a constant
of motion of the normal form. Choosing initial conditions such that $J_x(0)=0$ freezes the dynamics to the center manifold up to order $N=2$. 
Motion on the center manifold is then described by the
2-DOF Hamiltonian:
\beq{nfcm}
K^{(CM)}(J_y,J_z,\theta_y,\theta_z)= \omega_y J_y + \omega_z J_z +K_2(J_y,J_z,\theta_y-\theta_z) \ .
\eeq
The linear part is just $H_0$ with $J_x=0$. The nonlinear part $K_2$ depends not only on the actions
but contains resonant terms depending on
the angles in the combination 
\beq{res_comb}
\psi \doteq \theta_y-\theta_z
\eeq 
corresponding to the 1:1 resonance. Therefore, we have 
$$\frac{d}{dt} \left( J_y+J_z \right)=0$$
along solutions and we check that the third integral of motion 
$$J_y+ J_z = i( Q_2P_2 +  Q_3P_3 )$$ 
coincides with the one introduced in \eqref{sec_int} whose value is denoted by $\E$.  The 
explicit expression of $K_2$ for the problem at hand is 
\beq{nfcm2}
K_2=
\alpha J_y^2 + \beta J_z^2+J_yI_z(\sigma+2 \tau \cos(2\psi))
\eeq
where the coefficients $\alpha$, $\beta$, $\sigma$, $\tau$, depending only on $\mu$, are explicitly given as series expansions in the boxes of \citet{CCP}.

\section{Geometric reduction}
\label{GRS}

In general terms, to compute a 1:1 resonant normal form around an equilibrium is equivalent to construct a polynomial series invariant under the 
Hamiltonian flow of the planar isotropic oscillator. This dynamical symmetry is usually referred to as the $S^1$-action:
\beqa{HO_symmetry}
\Phi &:& S^1\times \RN^4 \longrightarrow \RN^4 \\
&& (\phi;P_2,P_3,Q_2,Q_3)\mapsto(P_2 e^{i \phi},P_3 e^{i \phi},Q_2 e^{-i \phi},Q_3 e^{-i \phi}) \label{HO_symm}\ .
\eeqa
The angle $\phi$ is an arbitrary phase along the closed orbits of the linear oscillator. The normal form \eqref{nfcm}, when expressed in the coordinates $(P_2,P_3,Q_2,Q_3)$, is an  example of this class of Hamiltonians. A general result concerning polynomials invariant under the action of a compact group is the Hilbert-Schwartz theorem \citep{HH18}: it states that a finite number of polynomials, generating the {\it Hilbert basis}, exist such that any invariant polynomial can be expressed in terms of them. 

\citet{CB} prove that the Hilbert basis of the polynomials in $\RN^4$ invariant under the $S^1$-action \eqref{HO_symmetry} is given by
\beqa{inv_basis}
I_0 &=&  i( Q_2P_2 +  Q_3P_3 ) = J_y + J_z = \E, \label{inv1}\\
I_1 &=&  i( Q_2P_3 +  Q_3P_2 ) = 2 \sqrt{J_y J_z} \cos\psi, \label{inv2}\\
I_2 &=&  Q_2P_3 -  Q_3P_2 = 2 \sqrt{J_y J_z} \sin\psi, \label{inv3}\\
I_3 &=&  i( Q_2P_2 -  Q_3P_3 ) = J_y - J_z \ . \label{inv4}
\eeqa
In each line of the above equations, the invariants are written both as quadratic polynomials (the so called {\it Hopf variables}) and in action-angles variables, the {\it Lissajous-Deprit} form \citep{D1,DE,Lara17}. The members of the Hilbert basis $\{ I_0,I_1,I_2,I_3\}$ form a Poisson algebra under the Poisson brackets given by 
\beqa{inv_basis}
\{I_j,I_k\} &=& 2\e_{jk\ell} I_{\ell}, \; j,k,\ell=1,2,3, \\
\{I_k,I_0\} &=& 0, \ \forall \ k \ .
\eeqa
The integral $I_0$, coinciding with \eqref{sec_int}, is a {\em Casimir} of the algebra. 

There is a difficulty with a Hilbert basis: not every generator is independent and this implies that the expression of a given invariant polynomial is not unique. In our case, it is easy to check that the members of the Hilbert basis (\ref{inv1}--\ref{inv4}) are subject to the relation
\begin{equation} \label{rel_invariants}
I_1^2+I_2^2+I_3^2=I_0^2 \ .
\end{equation}
However, a splitting of the possible expressions in which the Casimir $I_0$ is factored out seems to be a natural choice and this usually removes the non-uniqueness of the expression. On the other hand, the invariant relation \eqref{rel_invariants} determines, for each fixed $I_0=\E$, a 2-sphere in $\RN^3$ 
\begin{equation}\label{phase_sphere}
\mathcal S=\left\{(I_1,I_2,I_3)\in\mathbb R^3\;:\;I_1^2+I_2^2+I_3^2=\E^2\right\}
\end{equation}
to which we refer as the {\em reduced phase space}. Dynamics in terms of the Poisson variables $\{I_1,I_2,I_3\}$ are called {\em reduced dynamics}. In the reduced dynamics, points on the sphere correspond to $S^1$-orbits in  $\RN^4$ and, since the flow of the normal form preserves $I_0$, the reduced flow is simply given by the intersection of the level sets of the Hamiltonian $K^{(CM)}$ with the 2-sphere determined by \eqref{rel_invariants}. In practice, the orbits of the reduced flow are given by the curves provided by the two equations
\beqa{hamiltonian_inv}
K_I(I_1,I_2,I_3; \E)&=&{\mathcal K}_I \ , \label{K1}\\
I_1^2+I_2^2+I_3^2 &=& \E^2 \ , \label{S2}
\eeqa
where we have introduced the rescaled Hamiltonian 
\begin{equation} \label{KK_invariants}
K_I  (I_1,I_2,I_3; \E) \doteq \frac1{\omega_z} \Pi \circ K^{(CM)} = \left(1+\Delta\right)\E + A_1 \E^2 +(B\E+\Delta)I_3 + A I_3^2 + C (I_1^2-I_2^2) \ ,
\end{equation}
in which the projection map defined in  (\ref{inv1}--\ref{inv4}) 
\beqa{PRO_symmetry}
\Pi &:& \RN^4 \longrightarrow \RN^4 \\
&& (P_2,P_3,Q_2,Q_3)\mapsto( I_0,I_1,I_2,I_3)
\eeqa
and the shorthand notation
\beq{ABC}
A\doteq\frac{\alpha+\beta - \sigma}{4\omega_z},\;\;\;
A_1\doteq\frac{\alpha+\beta + \sigma}{4\omega_z},\;\;\;
     B\doteq\frac{\alpha-\beta}{2\omega_z},\;\;\;
C\doteq\frac {\tau}{2\omega_z},\;\;\; \Delta\doteq\frac{\d}{2\omega_z}
\eeq
have been used. The geometry of the level sets $K_I={\mathcal K}_I$ of the function $K_I(I_1,I_2,I_3; \E)$ on the 2-sphere $I_0=\E$ provides all informations on the global structure of the center manifold.

\section{Further reduction and analysis}
\label{FGRS}

We perform a further reduction introduced by \citet{HS} to explicitly divide out the additional symmetry of our system. In fact, we see that the two reflections $\mathcal S_2$ and $\mathcal S_3$ of \eqref{spatial_symmetry} now turn into the discrete symmetry $(I_1,I_2,I_3)\rightarrow(-I_1,-I_2,I_3)$ of \eqref{KK_invariants}. The reduction is then provided by the transformation
\beqa{tr_lem1}
\T &:& \RN^3 \longrightarrow \RN^3 \\
&& (I_1,I_2,I_3)\mapsto(X,Y,Z) \ ,
\eeqa
with
\begin{equation}\label{tr_lem2}
\left\{
  \begin{array}{ll}
    X= I_1^2-I_2^2 \ , \\
    Y= 2I_1I_2 \ , \\
    Z= I_3 \ .
  \end{array}
\right.
\end{equation}
This turns the sphere \eqref{S2} into the `lemon' space
\begin{equation}\label{lemon}
\mathcal L=\left\{ (X,Y,Z)\in\mathbb R^3\;:\;X^2+Y^2=\left(\E+Z\right)^2\left(\E-Z\right)^2; \;\; |Z| \le \E \right\} \ 
\end{equation}
and the Hamiltonian \eqref{KK_invariants} into 
\begin{equation}\label{hamiltonian_lem}
K_R(Z,X)=\T \circ K_I = \left(1+\Delta\right)\E + A_1 \E^2 + C X+(B\E+\Delta)Z + A Z^2 \ .
\end{equation}
We observe that $K_R$ does not depend on $Y$: the level sets $K_R(Z,X) ={\mathcal K}_I$ are parabolic cylinders in the $ (Z,X,Y) $-space. The study of the reduced dynamics is therefore further simplified: rather than investigating the intersection of two surfaces, in view of the translation invariance of $K_R$, we can just study the intersection of the parabola 
\begin{equation}\label{parabola}
X(Z)=\frac1{C} \left(h -(B\E+\Delta)Z-A Z^2 \right) \ ,
\end{equation}
where, omitting constant terms, we have redefined the `energy' as
\begin{equation}\label{eneconv}
h \doteq {\mathcal K}_I - \left(1+\Delta\right)\E - A_1 \E^2 
\end{equation} 
and the contour $\mathcal C=\mathcal C_-\cup\mathcal C_+\doteq\mathcal L\cap\{Y=0\}$ in the $(Z,X)$-plane composed of the two parabolic segments
\be\label{lemon_arcs}
\C_\pm\equiv\left\{(Z,X)\in\mathbb R^2\;:\;|Z|\le\E,\;\,X=\pm\left(\E^2-Z^2\right)\right\} \ .
\ee
The price we pay for this second reduction is that {\em pairs} of critical points associated to periodic orbits in `generic position', respectively
\beqa{critp}
{\rm Halo}	\quad (\psi = \pm \pi/2) &:& \quad I_1 = 0, \quad I_2 = 	\pm \sqrt{\E^2 - Z_{-}^2}, \quad I_3 = Z_{-} \ ; \label{Ia}\\
{\rm Anti-halo}	\quad (\psi = 0, \pi       ) &:& \quad I_2 = 0, \quad I_1 = 		\pm \sqrt{\E^2 - Z_{+}^2}, \quad I_3 = Z_{+} \ ; \label{La}
\eeqa
which are distinct on the sphere $\mathcal S$, are instead represented by the {\em unique} tangency of the parabola \eqref{parabola} with either $\mathcal C_+$ (in $Z=Z_{+}, X(Z_{+})$) or with $\mathcal C_-$ (in $Z=Z_{-}, X(Z_{-})$). However, this indeterminacy does not hinder the correct description of the bifurcation scenario. In fact, by investigating when the tangencies
\begin{equation}\label{sl_int}
\left\{
  \begin{array}{ll}
  X= &X(Z)\\
  X\in &\C_{\pm}
  \end{array}
\right.
\end{equation}
exist for $|Z|\le\E$, namely when the discriminant of the quadric
\be\label{ZQZ}
X(Z)=\pm \left(\E^2-Z^2\right)\ee
vanish (see below the values in equations \eqref{REQ}), we get the thresholds for the bifurcations of periodic orbits in generic position from/to the normal modes $Z=\pm\E$. It is possible to check \citep{MP16} that, at
\beq{EL}
\E^-_{p,v}\doteq \frac{\Delta}{\pm2(A+C)-B} 
\eeq
the halo family bifurcate and at 
\beq{EU}
\E^+_{p,v}\doteq \frac{\Delta}{\pm2(A-C)-B} 
\eeq
the bridge family\footnote{This family is also dubbed `sideway' family \citep{FJM}} bifurcate from the planar and annihilate on the vertical Lyapunov: here $-,+$ denote the two families and $p,v$ denote the `planar' and `vertical' normal modes. We observe that the annihilation predicted by \eqref{EL} at $\E^-_{v}$ is actually impossible because its value is always negative in these systems.

\begin{figure}[h]
\center
\includegraphics[height=6cm]{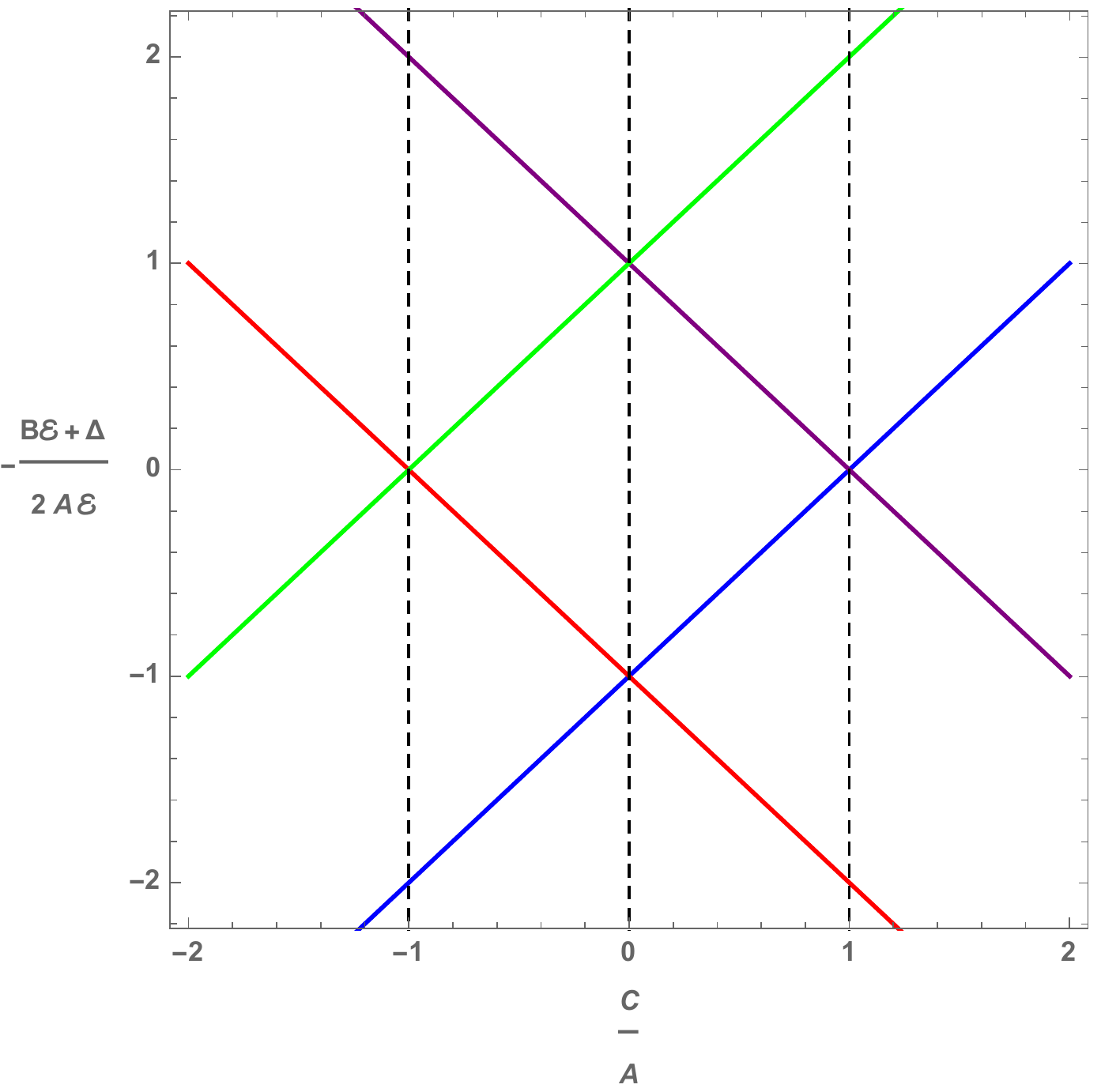}
\includegraphics[height=6cm]{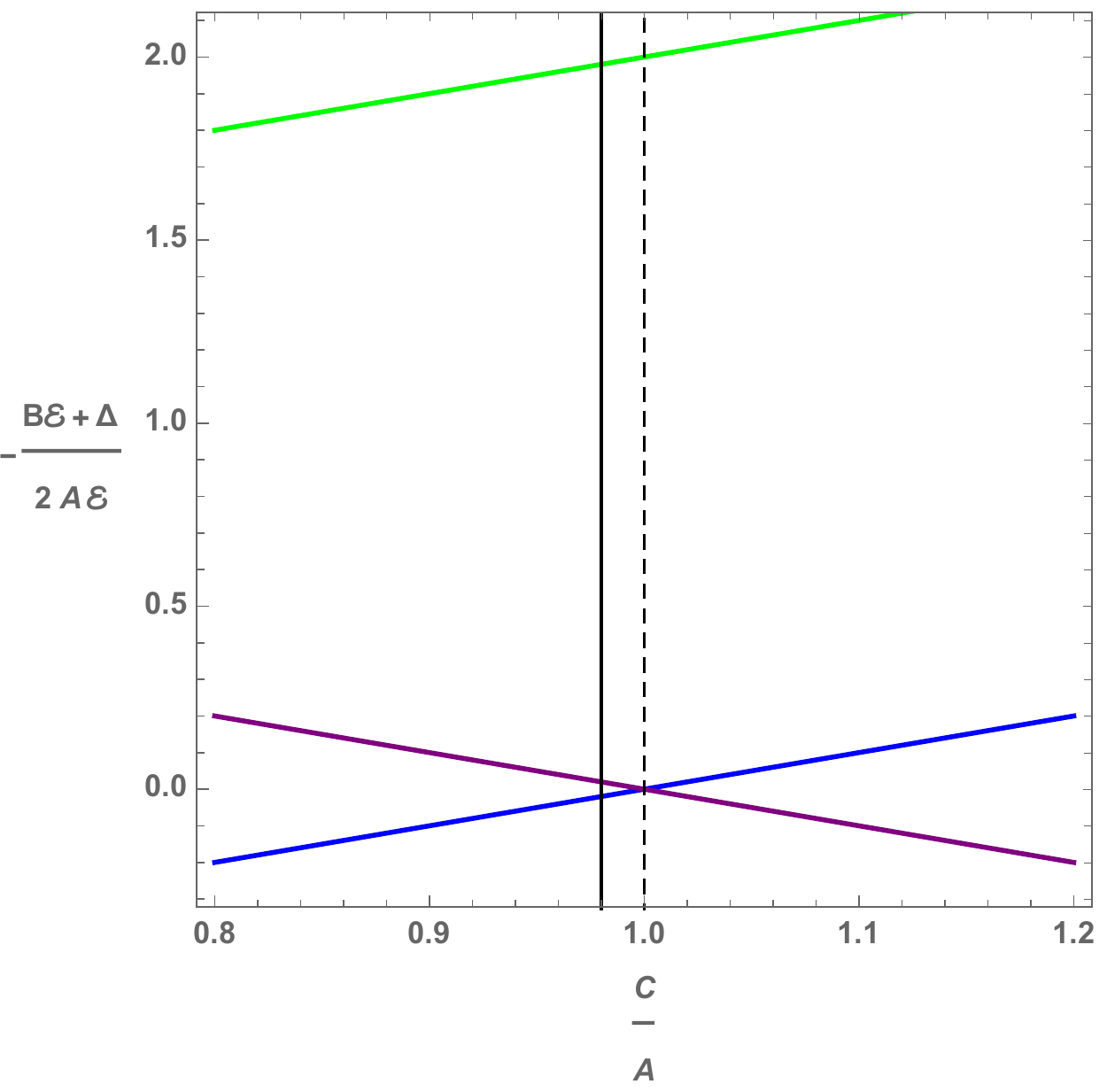}\\
\caption{Left: the bifurcation lines associated with $\E^-_{p,v}$ (eq.\eqref{EL}, green and red lines), $\E^+_{p,v}$ (eq.\eqref{EU}, blue and purple lines) and vertical dashed lines denoting the degenerate cases $|C/A|=0,1$. Right: enlargement with the continuous vertical line followed by the system with $\mu=1/2$ around $L_2$.
}\label{F1}
\end{figure}

An additional important advantage of the second reduction concerns generic properties of our family of models. When in \eqref{parabola} the ratio $|C/A|$ is equal to one, the curvature of the parabola $X(Z)$ is the same as that of the parabolic segments $\mathcal C_{\pm}$ and, instead of having critical points, we have {\em critical circles}. Therefore, we get an easy way to identify the non-generic case (typical of what happens with the H\`enon-Heiles system \citep{HS}) which requires to go to a higher order in the normalisation to disentangle the role of perturbations. However, \eqref{parabola} is characterised by four independent parameters, $A,B,C,\Delta$ which is still too much for a clear global representation. In \citet{MP16} a representation in terms of only two independent parameters have been introduced which allows us an easy interpretation of the general setting. The two parameters are just the curvature (or `coupling') parameter $C/A$ and the `asymmetry' parameter
\begin{equation}\label{zmp}
\frac{Z_V(\E)}{\E}=-\frac{B\E+\Delta}{2A\E} \ .
\end{equation}
In the $(Z,X)$-plane, $Z_V$ denotes the abscissa of the vertex of the parabola. Substituting the threshold values (\ref{EL},\ref{EU}) into \eqref{zmp} we get the four lines 
\begin{equation}\label{zmpb}
\frac{Z_V(\E^-_{p,v})}{\E^-_{p,v}}=\pm \left(1 + \frac{C}{A}\right), \quad  \frac{Z_V(\E^+_{p,v})}{\E^+_{p,v}}=\pm \left(1 - \frac{C}{A}\right) \ .
\end{equation}
In the parameter plane, a given system (characterised by fixed values of the parameters) `moves' along a vertical straight line as $\E$ varies. When this line crosses one of the lines in the set \eqref{zmpb}, a bifurcation occurs. This structure is generic  in every point of the plane with the exception of  the degenerate cases $|C/A|=0,1$ (see the left panel in Fig. \ref{F1}). In any point in the plane with $|C/A|\ne0,1$ (and not on a bifurcation line), any sufficiently small change of the parameters does not modify the phase space structure. Among possible changes we may include `perturbations' consisting of higher-order terms of the normal form. Higher-order normalisation is therefore needed only in order to get more accurate quantitative predictions.

\begin{figure}[h]
\center
\includegraphics[height=7cm]{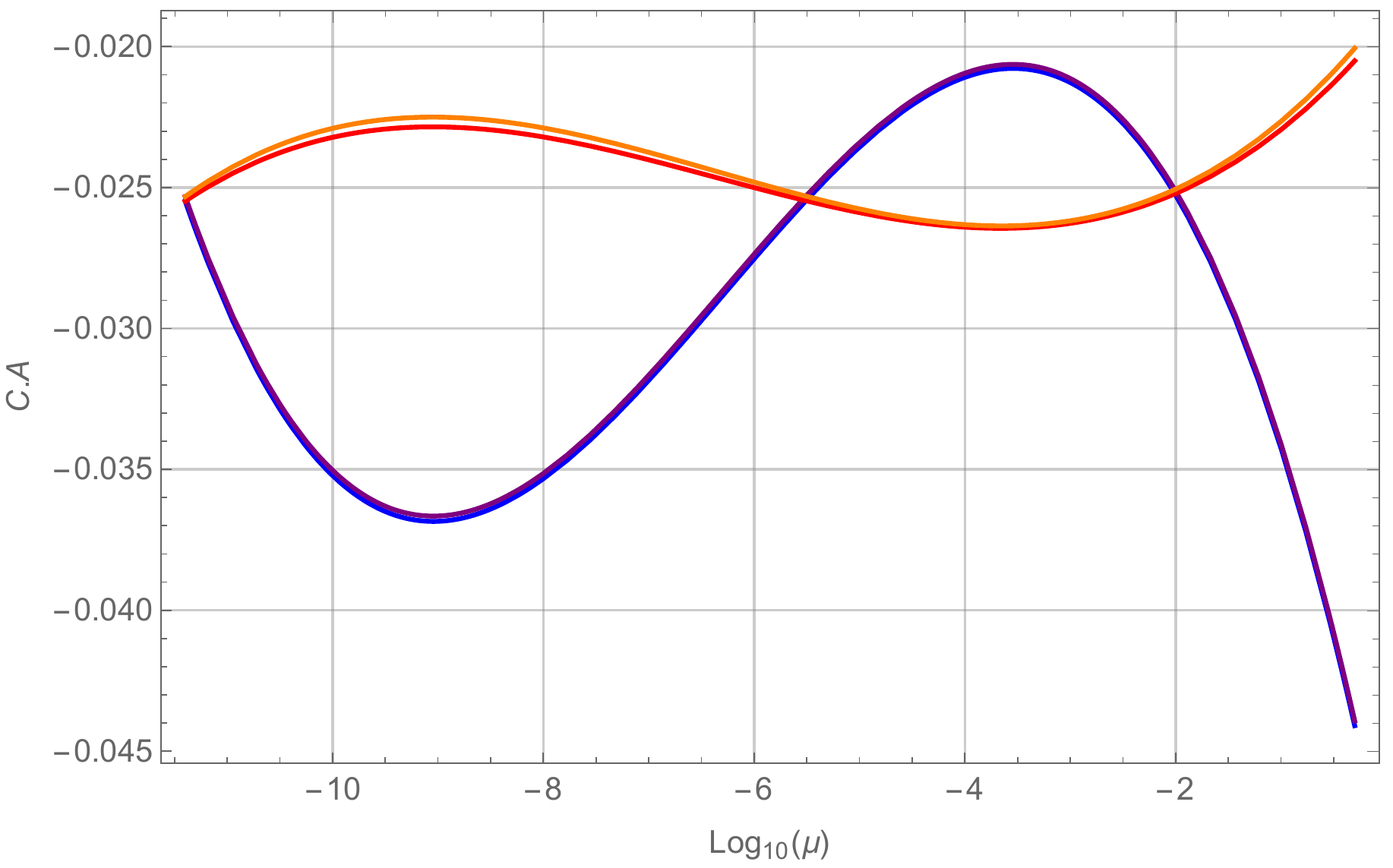}
\includegraphics[height=7cm]{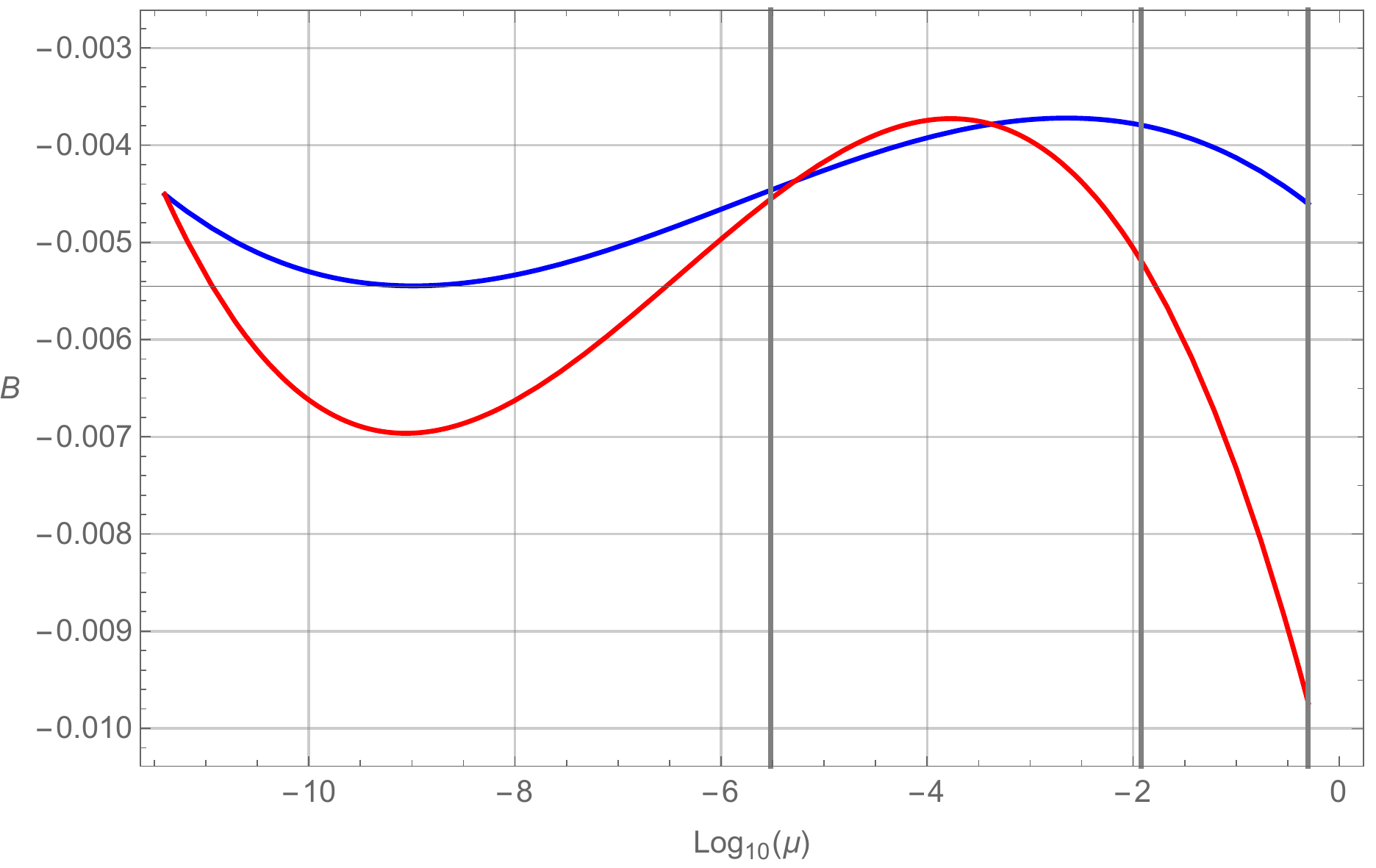}
\caption{ Coefficients of the reduced normal form \eqref{hamiltonian_lem} as functions of the mass ratio. Upper panel: for $L_1$, the curves for $A(\mu)$ (blue) and $C(\mu)$ (purple) are almost overlying; for $L_2$, $A(\mu)$ (red) and $C(\mu)$ (orange) are more distinguishable. Lower panel: $B(\mu)$, (blue for $L_1$) and (red for $L_2$). Grey lines again correspond to the three `reference' cases: Earth-Sun, Earth-Moon and equal masses. 
}\label{F2A}
\end{figure}

\begin{figure}[h]
\center
\includegraphics[height=7cm]{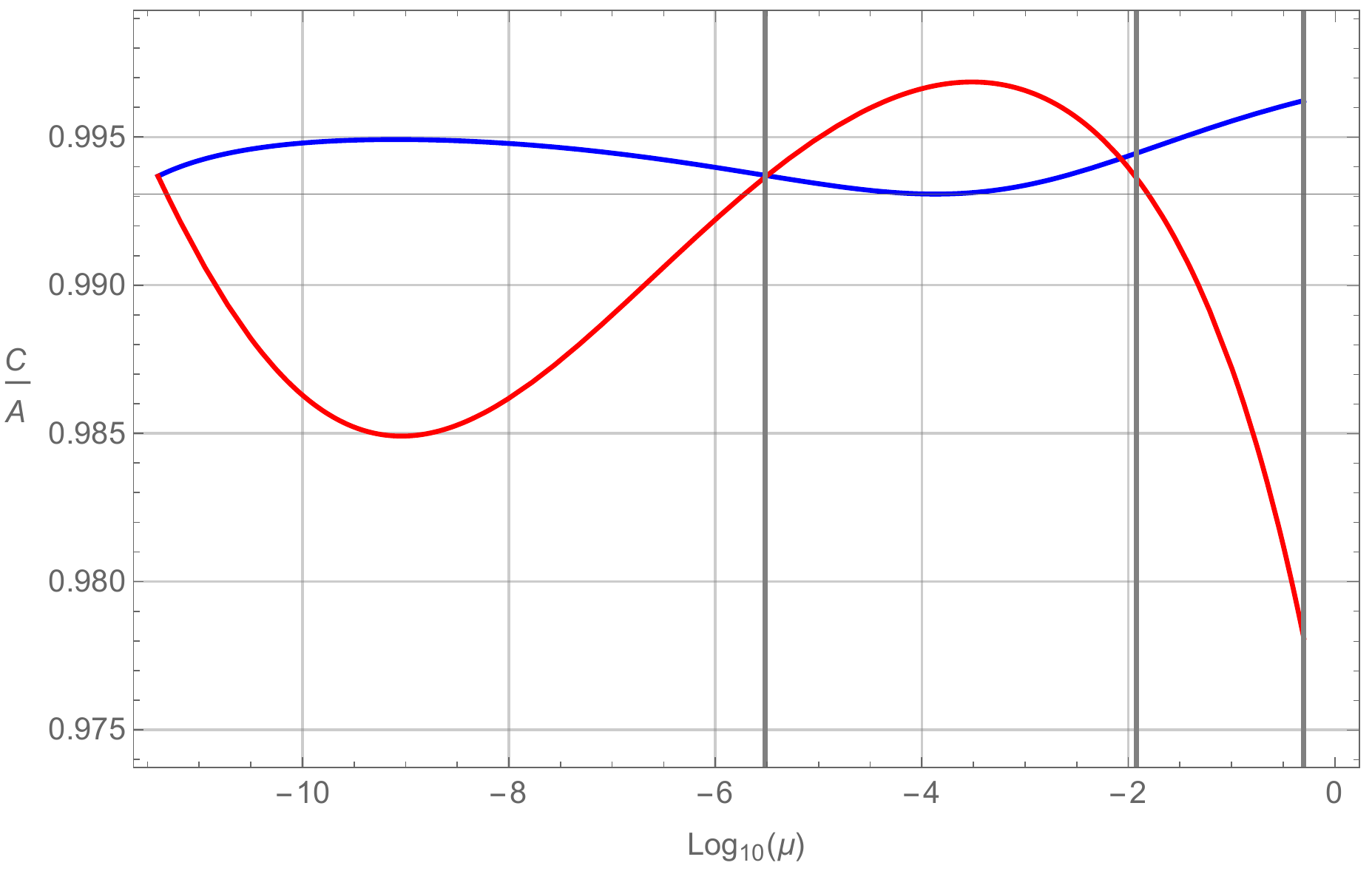}
\caption{Coefficient ratios $C/A$ as functions of the mass ratio for $L_1$ (blue curve) and $L_2$ (red curve).
}\label{F2B}
\end{figure}

Let us see how all this applies to the case at hand. The parameters $A,B,C,\Delta$ for the center manifolds of the collinear points can be computed starting from the values found in \citep{CPS} and \citep{CCP}. In this last paper, explicit series expansions for the coefficients of the normal form are given. It happens that, for every $\mu$ in the range $(0,1/2]$, the coupling parameter $C/A$ is always slightly less than $+1$: this is shown in  the upper panel of Fig. \ref{F2A} and, with more detail, in Fig. \ref{F2B}.  Choosing the case with $\mu=1/2$ around $L_2$ which has the minimum value $C/A=0.978$, we plot its `evolution line' (continuous) aside the degenerate line (dashed) in the right panel in Fig. \ref{F1}. Note that, with the present values of $A,B$ and $\Delta$, energy increases along the evolution line going from top to bottom, so that the halo bifurcation (crossing the green line) first occurs and then the bridge of subsequent anti-halo bifurcations (crossing purple-blue lines) appears. All other cases with mass parameter in the range $(0,1/2]$ share the same picture.

The overall phase-space structure can be assessed by plotting the `energy-momentum' map \citep{Arn1,CB}. In the present case, let us again split the rescaled normal form in the center manifold as 
\beq{nfcms}
\widetilde K^{(CM)}(P_2,P_3,Q_2,Q_3)= \widetilde K_0 + \widetilde K_2 \ ,
\eeq
where $\widetilde K_0 = \E$ is the isotropic quadratic part and $\widetilde K_2$ can be taken as the second integral. 
The energy-momentum map is then defined as the map
\beqa{EMAP}
\E{\cal M}: && \mathbb R^4 \longrightarrow  \mathbb R^2\\
&& w \longmapsto \left(\widetilde K_0 (w), \widetilde K (w)\right),\\
&& (P_2,P_3,Q_2,Q_3) \longmapsto  \left(\widetilde K_0 (P_2,P_3,Q_2,Q_3), \widetilde K (P_2,P_3,Q_2,Q_3)\right)\eeqa
where we use, as second component, $\widetilde K=\widetilde K_2-\Delta\E - A_1 \E^2=h$, in agreement with
\eqref{eneconv}. Points $w \in \mathbb R^4$ are called {\it regular} if the rank of the matrix of the differential of the map
\be
d{\cal EM}=\begin{pmatrix}
\partial_w \widetilde K_0 \\
\partial_w \widetilde K \\
\end{pmatrix}
\ee
is maximal: two, in the present case. Otherwise, if ${\rm rank} [d{\cal EM} (w_c)] < 2$, $w_c$ are called {\em critical points} and ${\cal EM}(w_c)$ are the {\em singular level sets}. 

The information provided by the singular sets in the image of the ${\cal EM}$-map complement those offered from the Liouville-Arnol'd theorem. In the present context this theorem proves that, chosen a regular value of the map, there is a neighbourhood $W(w) \subset \mathbb R^2$ such that ${\cal EM}^{-1} (W)$ is foliated by invariant tori. However, there is no such general statement concerning the topology of the inverse image of the singular level sets: they determine a bifurcation diagram that, in general, depends on the system at hand and gives the `skeleton' of the phase-space by organising the domains where different families of ordinary 2-tori exist. 

The singular sets are in the form of `branches' whose inverse image, in the present 2DOF case, are either stable periodic orbits or unstable periodic orbits together with their stable/unstable manifolds. These branches separate connected components of the domains of the map containing regular values corresponding to 2-tori. To plot the branches we need to find the critical values 
$$\left(\widetilde K_0 (Z_c,X_c), \widetilde K (Z_c,X_c)\right) \ ,$$ 
where $(Z_c(P,Q),X_c(P,Q))$ collectively denote critical points. They are given by the singular equilibria associated to the normal modes $Z=\pm\E,X=0$,
\beqa{SEQ}
\widetilde K_0 = \E \ , \quad
\widetilde K = \E \left((A \pm B) \E - \Delta \right)
\eeqa
and by the tangency conditions \eqref{sl_int} of the energy surface with the reduced phase space given by
\beqa{REQ}
\widetilde K_0 = \E  \ , \quad
\widetilde K = \pm C\E^2-\frac{(B\E+\Delta)^2}{4(A \mp C)} \ .
\eeqa
These curves are parabolic arcs respectively representing the values of the two components of the ${\cal EM}$-map on the normal modes (the two curves \eqref{SEQ}) and the values on the regular equilibria $Z_{\pm}, X(Z_{\pm})$  (the two curves \eqref{REQ}).

\begin{figure}[h]
\center
\includegraphics[height=6cm]{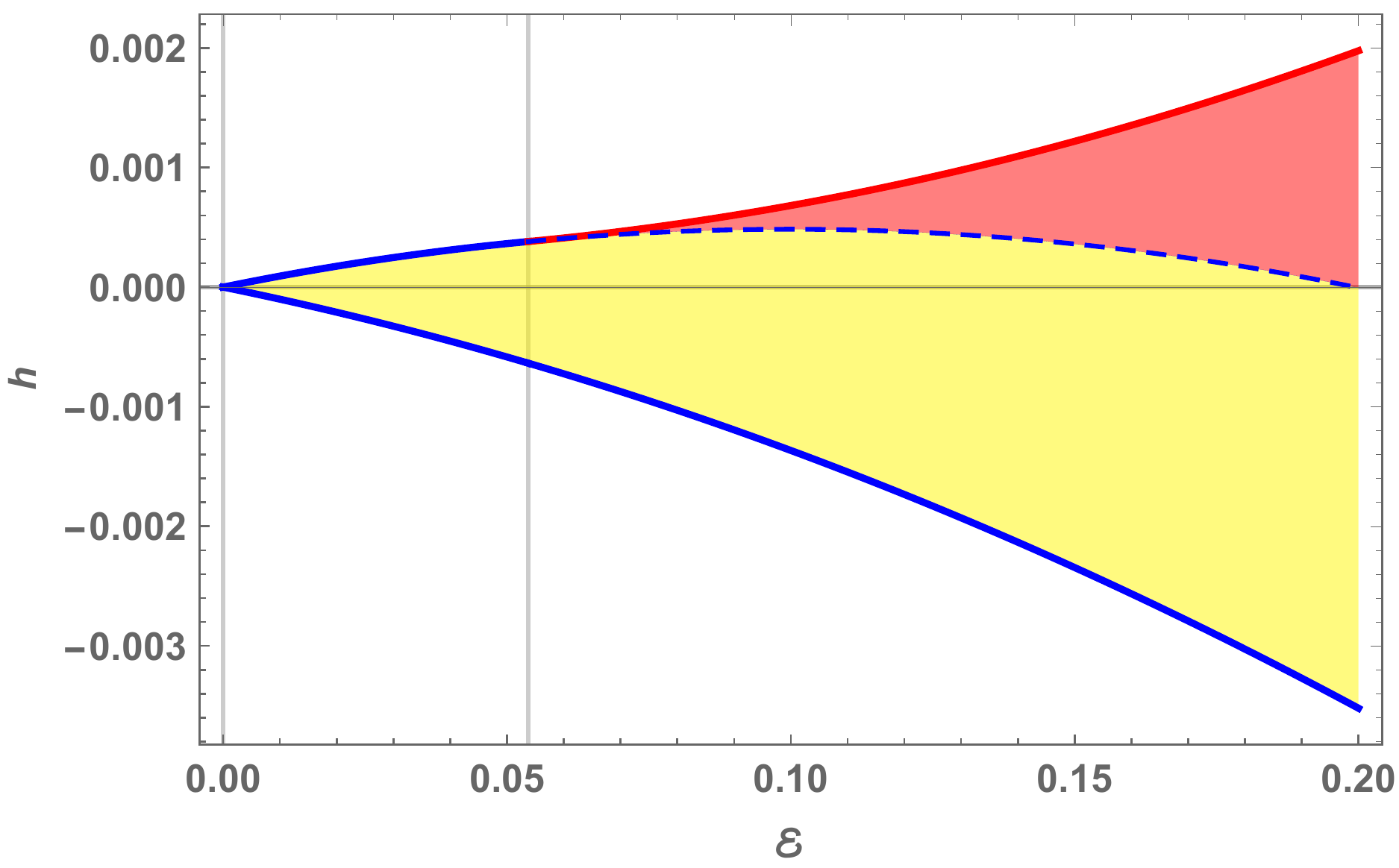}\\
\caption{Energy-momentum map for the system with $\mu=1/2$ around $L_1$: the bifurcation of halo orbits.
}\label{F3}
\end{figure}

In Fig. \ref{F3} we plot the domains in the $(\E,h)$-plane around the first bifurcation occurring in the system with $\mu=1/2$ at $L_1$. The blue branches are the arcs \eqref{SEQ} and represent the Lyapunov orbits: the lower one is the vertical family and the upper one is the planar family. At $\E^-_{p} = 0.054$, obtained using \eqref{EL} with the data of Figg. \ref{F2A},\ref{F2B}, the halo family (in red, the curve in eq.\eqref{REQ} with the lower sign) bifurcates from the planar Lyapunov\footnote{We observe that $E_p = \omega_z  \E^-_{p} = 0.153$ is therefore the analytic first-order prediction of the bifurcation energy of the halo orbits around $L_1$ in the equal-masses case. The numerical value obtained by \citet{Henon1} was $E=0.153860$.} which passes from stable to unstable (dashed line). The yellow domain represents invariant tori around the stable vertical, the red one invariant tori around the stable halo.  The corresponding plot for $L_2$ is topologically the same but for the bifurcation value $\E^-_{p} = 0.334$. 

\begin{figure}[h]
\center
\includegraphics[height=6.4cm]{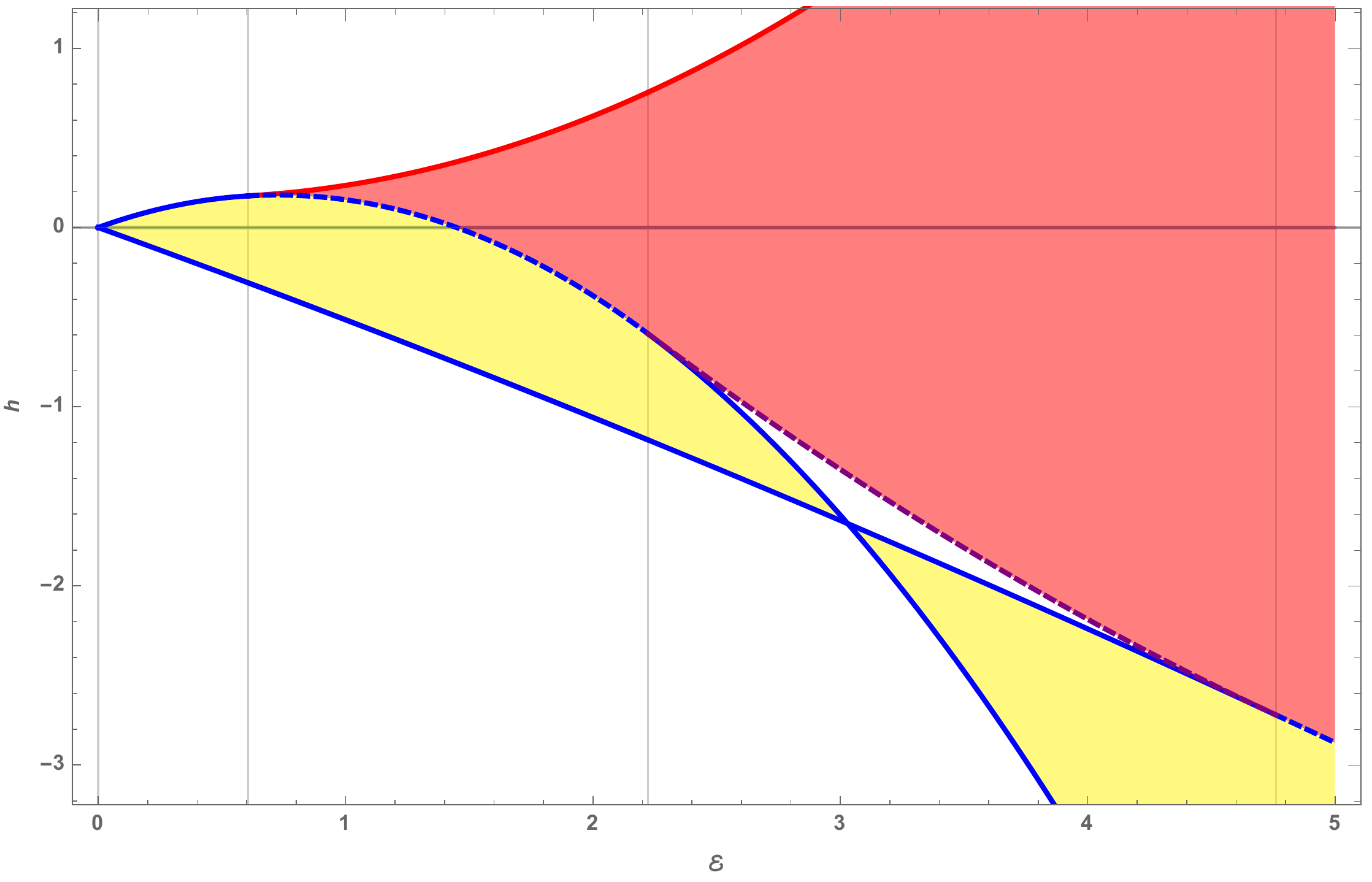}\\
\includegraphics[height=4.8cm]{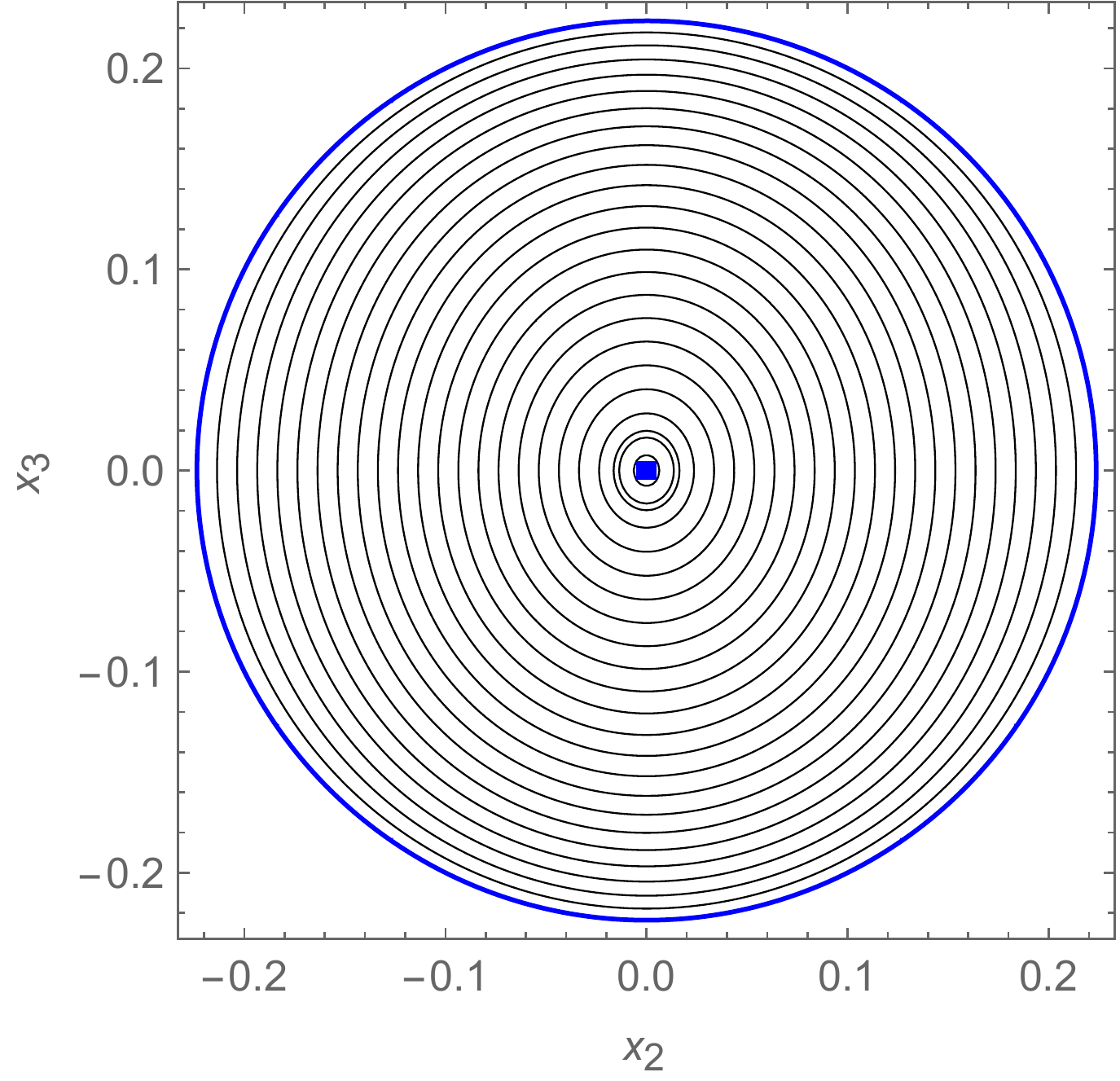}
\includegraphics[height=4.8cm]{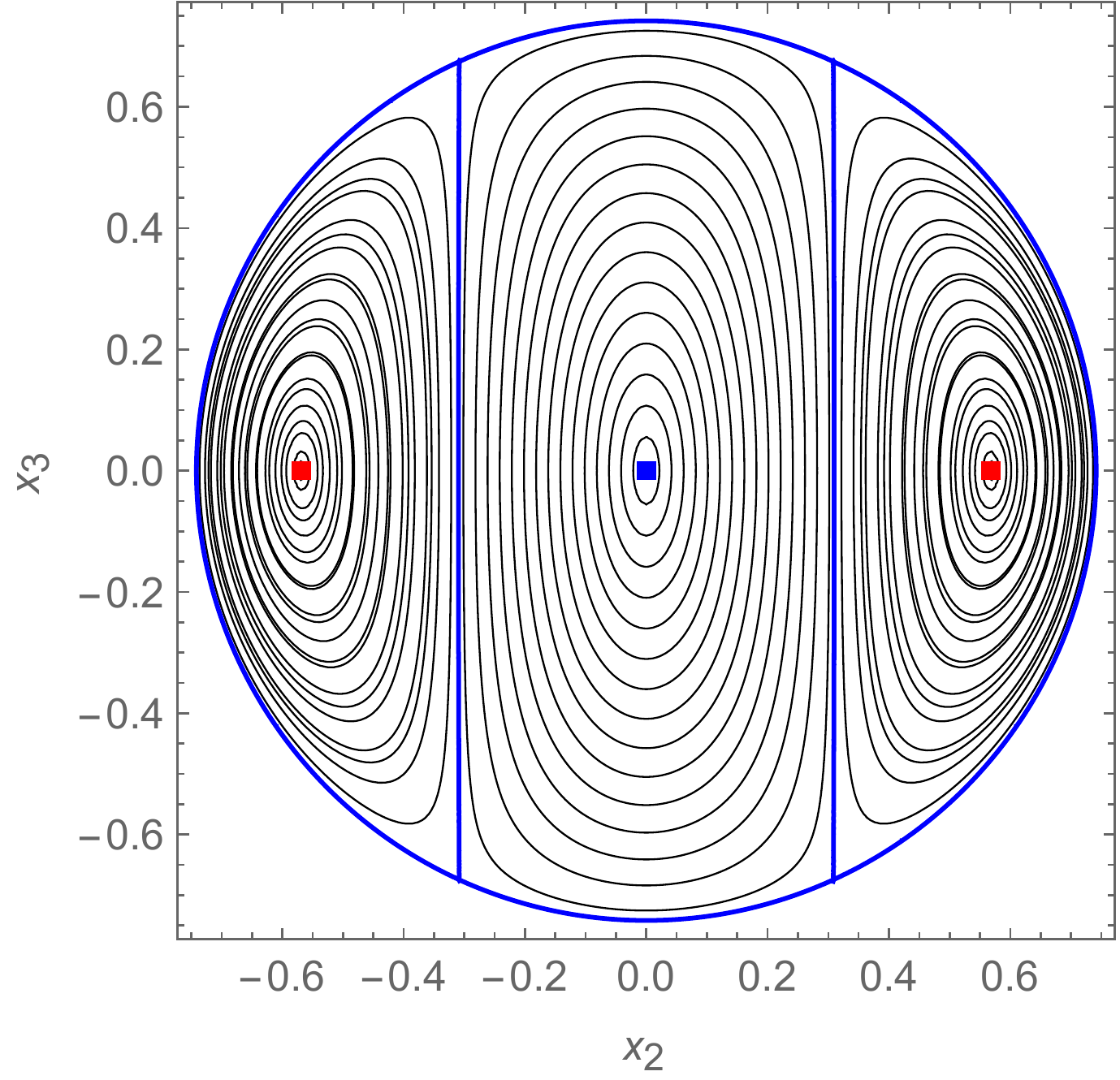}\\
\includegraphics[height=4.8cm]{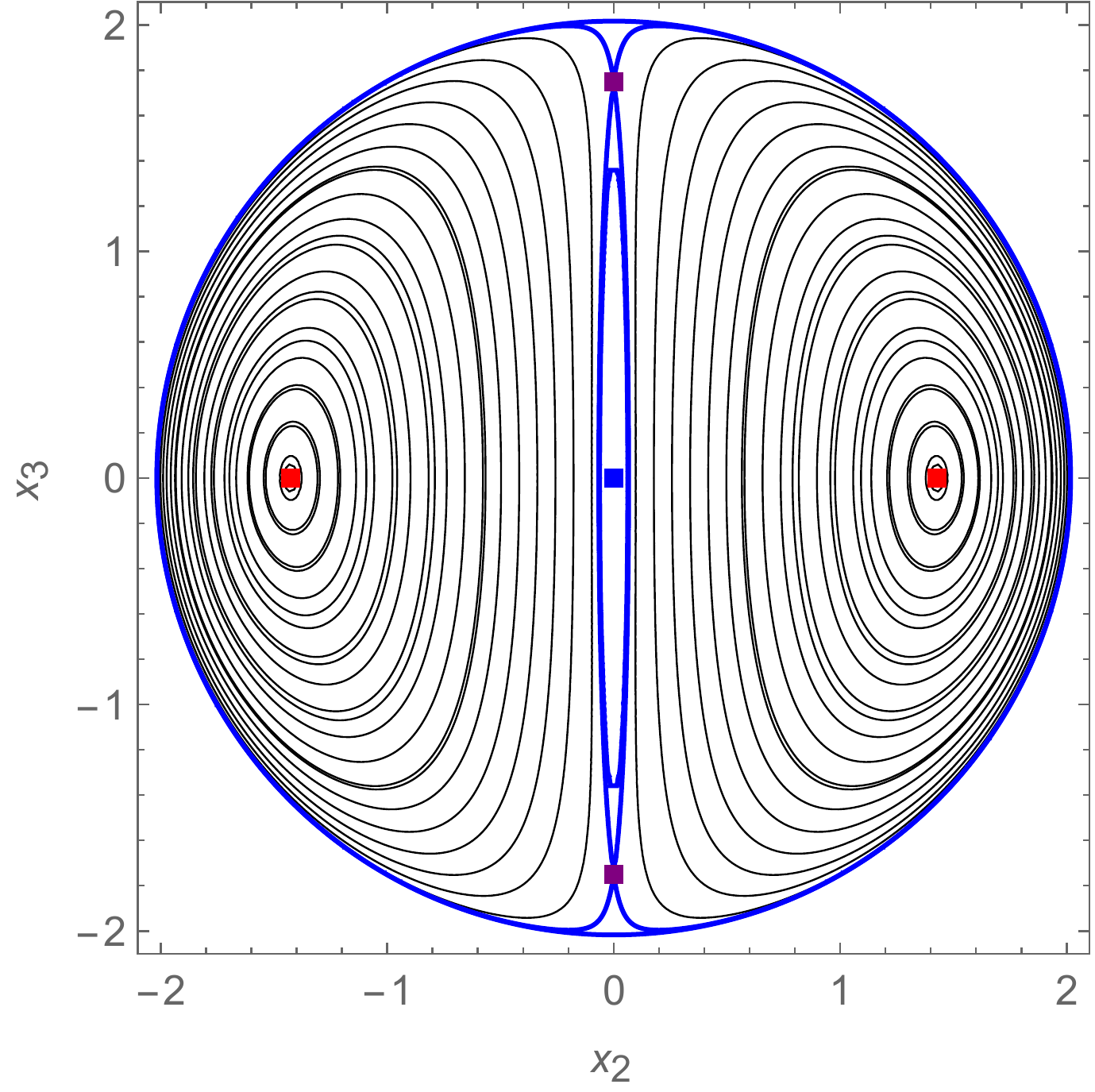}
\includegraphics[height=4.8cm]{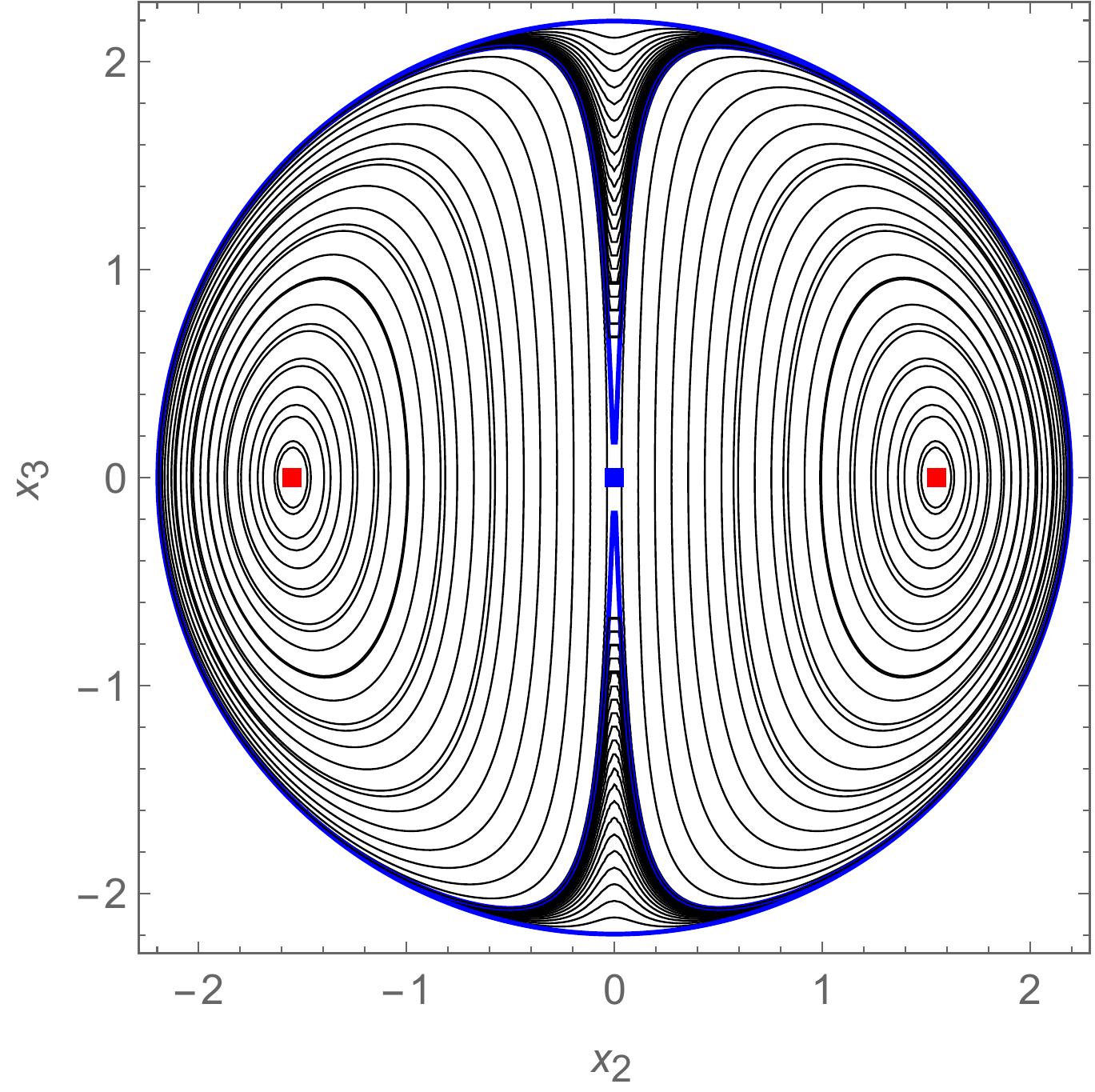}\\
\caption{Upper panel: energy-momentum map showing the bifurcation set for the system with $\mu=1/2$ around $L_1$. Lower panels: analytic Poincar\'e sections with the possible inequivalent phase flows. 
}\label{F4}
\end{figure}

Fig. \ref{F3} is an enlargement around the origin of Fig. \ref{F4}. In Fig. \ref{F4} we plot the $(\E,h)$-plane including the bifurcation and collision of the bridge sequence: at $\E^+_{p} = 2.22$ the (unstable, dashed) anti-halo family bifurcates from the planar Lyapunov which regains stability; at $\E^+_{v} = 4.76$ the anti-halo family disappears colliding with the vertical Lyapunov which loses stability. Yellow domains still represent tori around stable normal modes, the red one tori around the stable halo and the white domain the intermediate family bounded by the separatrix composed of the stable/unstable manifolds of the unstable bridge orbits. 

Clearly, the bifurcation energies predicted by the first-order theory are poor predictors of the true values obtained with numerical integrations \citep{Henon1,GM}, for more accurate predictions higher-order terms are necessary. However, for the sake of the present analysis the overall picture is correctly reconstructed. The Poincar\'e surfaces of sections shown in the lower panels of Fig. \ref{F4} are computed by plotting, in the plane
\beqa{PPSS}
x_2= \sqrt{2} \Re\{P_2\} = - \sqrt{2} \Im\{Q_2\}, \quad x_3 = \sqrt{2} \Re\{P_3\} = - \sqrt{2} \Im\{Q_3\} \ ,\eeqa
the level curves of the function
\beqa{PSS}
F(x_2,x_3;\E)=K^{(CM)}(P_2,P_3,Q_2,Q_3) \bigg\vert_{\Im\{P_3\}=0, \Re\{Q_3\}=0} 
\eeqa
with the energy constraint
\beqa{EC}
\Re\{Q_2\}= - \Im\{P_2\} = \sqrt{\E-\Re\{P_2\}^2-\Re\{P_3\}^2} \ .
\eeqa
The choice of the variables $x_2,x_3$ complies with that usually made in computing numerical surfaces of section \citep{JM,FJM}. In our case, still referring to the value $\mu=1/2$ around $L_1$, they are plotted at values of $\E$ below and above the halo bifurcation level $\E^-_{p}$ (upper panels), just above the bifurcation of the bridge ($\E^+_{p}$, lower left panel) and just above its annihilation ($\E^+_{v}$, lower right panel). Note that, by using the unreduced normal form, namely not implementing the second reduction (\ref{tr_lem1}--\ref{tr_lem2}), we correctly retrieve the pairs of fixed points corresponding to the {\em two} northern and southern haloes (red dots) and the double bridge of the anti-haloes (purple dots). The planar Lyapunov is given by the boundary of the section and the vertical Lyapunov by the central blue dot.

By inverting the normalising transformation we can express motion solutions in terms of the original variables $(p,q)$ and, after that, $(p_x,p_y,p_z,x,y,z)$. In particular, we can obtain approximate expressions for the periodic and quasi-periodic motion and other invariant objects and remove one of the mirror symmetries enforced by the normalisation. However, the explicit expression of the normal form Hamiltonian in the original coordinates forbids the simple geometric analysis developed here.

\section{Implications for the original system}

In view of the involved form of the diagonalising transformation, the investigation of the dynamics around the collinear points is usually performed by computing solutions for specific values of the mass parameter. In the literature is now available a good deal of results ranging from the Hill limit up to primaries with comparable size. We have presented a general approach to unify this body of knowledge in a comprehensive framework valid for arbitrary values of the mass parameter. The geometric picture describes all possible bifurcations related to the approximate isochronous resonance and provides accurate predictions for the halo orbits and a less accurate, but qualitatively correct, description of the bridge of anti-halo orbits. 

 To come back from the reduced one-degree-of-freedom system to the flow in two degrees of freedom (the `original system'), we first exploit the normal form symmetry embodied in \eqref{HO_symm}: we have to attach an $S^1$-circle to every point of the reduced phase space $\mathcal S$. This is most simply done in the Lissajous-Deprit variables of Section 3, where the action-angle pair $\E,  \theta_y+\theta_z$ parametrising the circles has a (fast) frequency of order 1 and then equilibria on $\mathcal S$ give rise to periodic orbits, while periodic orbits on $\mathcal S$ yield invariant 2-tori of the flow of $K^{(CM)}$. Second, given the normalisation procedure, we may consider $H$ as a perturbed form of $K$ with the remainder $R$ playing the role of perturbation. Iso-energetic KAM-theory \citep{Arn1} allows to conclude that most invariant 2-tori survive the perturbation if is sufficiently small; on each energy surface the complement of persisting 2-tori has a small measure. The iso-energetic non-degeneracy condition associated to the quadratic part of $K^{(CM)}$ depends on $A,B,\Delta$ and can be easily checked to be always ($\forall \mu \in (0,1/2]$) satisfied. Existence and stability (with related bifurcations) of periodic orbits in the original system are deduced by means of the implicit function theorem. 

 It is evident that if we desire better predictions for the occurrence, in the original system, of these bifurcation and other dynamical features like Lissajous quasi-periodic orbits, invariant tori around the halos, etc., a high-order normalisation is mandatory. However, in view of the asymptotic nature of the series so obtained, an \sl optimal \rm order is usually reached which prevent to get arbitrary accuracy. Rather, as a rule, the domain of validity of the asymptotic series may shrink with increasing order. A common feature of such divergent series is to belong to some Gevrey class \citep{GS}: for example, the higher-order series expansions of the bifurcation thresholds found in \citep{CCP} appear, from numerical evidence, to belong to the Gevrey-1 class.  Standard partial sums of terms obtained at each step of the normal form construction may therefore provide not very good results. We conjecture that re-summation techniques of non-convergent series may be profitable in this setting \citep{Sc1,PBB}. 

 Anyway, the normal form itself seems to have good semi-convergence properties: the sequence of partial sums shows a mild logarithmic divergence \citep{JM} and this allows to make estimates on the domain of validity of the approximation.  Its extent amply accommodates for the bifurcations investigated here. More delicate is the behaviour of the normal form in the case of $L_3$, where, at low mass ratios, the optimal order seems to be so low to render very uncertain the estimate of the domain. However, we remark that the first-order prediction for the bifurcation of the halo made in \citet{CCP} is in good agreement with recent accurate numerical estimates \citep{WW} also for $\mu \to 0$. 

In the light of the geometric picture described here, we may wonder why all systems are so close (but always bounded away) to the degenerate case in which the addition of a small perturbation leads to unpredictable bifurcation sequences. Just looking at the values of the coefficients it seems that this situation is more or less accidental. However, an interesting remark we can provide concerns `artificial' changes of the parameters. This is possible when considering the third body to be a solar sail subject to the solar radiation pressure (SRP, \citet{colin}). In the simple case in which the sail is always perpendicular to the light source, the problem is Hamiltonian and can be treated along the lines followed above \citep{BMC}: it can then be shown that, for physically acceptable values of the `performance parameter', the energy bifurcation thresholds are always {\em lowered} with respect to those of the purely gravitational case \citep{AMPA,FJM}. We can therefore state that the representative evolution line of systems with SRP are displaced to the left in a plot on the parameter plane like the one in the right panel of Fig. \ref{F1}: that is, halo and bridge orbits appear at lower energies and the bridge family annihilate at higher energies. The degenerate case $C/A=1$ can be reached only in the unphysical case of a negative performance parameter. However, we cannot exclude that other forms of perturbations may induce a change of this sort in the bifurcation scenario.  An interesting possibility is, for example, offered by considering the elliptic restricted problem. In this case, the collinear points are no more strictly equilibria. However, several bifurcation phenomena are still present \citep{HouLiu} and can be investigated by a suitable generalisation of the present approach. To take into account the non-autonomous character of the problem, the resonant normal form has to be computed in the extended phase-space. However, it should still be an integrable approximation with two parameters, $\mu$ and the eccentricity of the primaries. 

\section{ Hints for further study and  conclusions}

There are several hints for additional studies:  in addition to the already mentioned elliptic extension,  the analysis of bifurcations of families associated to other resonances can be performed by constructing appropriate resonant normal forms. Analogously, we can think to the investigation of secondary resonances, like those appearing at high energy around the halo in the computations of \citet{GM}. The geometric analysis of the 1:2 and 1:3 resonances proceed along lines quite similar to those followed above \citep{CDHS} and could offer a comprehensive view of orbit families departing from existing periodic orbits. For these orbits, like for the Lissajous and halo families, remains to be proven the complete equivalence between solutions computed with the Poincar\'e-Lindstedt method \citep{richardson,simo2,JM,LXC} and those found by inverting the normalising transformations. This equivalence is quite easily obtained in natural systems like the logarithmic potential  \citep{Sc2,PBB} but is still a hard task here in view of the cumbersome computations required by the coordinate transformations.

\section*{Conflict of interest}
The author declares that he has no conflict of interest.

\begin{acknowledgements}
We acknowledge useful discussions with A. Celletti, C. Efthymiopoulos, H. Han{\ss}mann, A. Giorgilli, M. Guzzo, A. Marchesiello and D. Wilkzak. The work is partially supported by INFN, Sezione di Roma Tor Vergata and by GNFM-INdAM.
\end{acknowledgements}

\end{document}